# Non-Resonant Below-Bandgap Two-Photon Absorption in Quantum Dot Solar Cells

Tian Li (李恬) and Mario Dagenais*

*Department of Electrical and Computer Engineering, University of Maryland, College Park, Maryland, 20742, USA*

**Abstract**: We are the first to show with clear experimental results that photons that have energy lower than the transition energy between quantum dots states and valance band can still contribute greatly to the photocurrent via both one-photon absorption process (1PA) and two-photon absorption process (2PA). The tailing states function as both the energy states for low energy photon absorption and the photocarriers extraction pathway. One of the biggest advantages of our method is that it can clearly differentiate the photocurrent due to 1PA process and 2PA process. Both 1PA and 2PA photocurrent generation efficiency in an InAs/GaAs QD device with photon excitation at 1550 nm have been quantitatively evaluated.

[*]Electronic mail: dage@ece.umd.edu.

Due to the 3-Dimentional quantum confinement of electrons in quantum dots, additional discrete energy levels within the bandgap are formed and potentially can function as intermediate states for additional sub-bandgap photon absorption. Recently we discovered that the strain at the interface of InAs and GaAs will create a large extended tailing distribution of density of states exponentially decaying into mid gap[1–3], which greatly affects the overall performance of solar cells. We believe that the existence of this tailing state is one of major obstacles among others in the efforts of realizing the concept of intermediate band solar cell model that was proposed by Marti and Luque[4,5]. In our previous paper, we have evaluated quantitatively the broadening of this extended tail and its contribution to the overall enhancement of Jsc[6]. A tail width of around 50 meV was obtained from an external quantum efficiency measurement compared with a 13 meV tail width for a bulk GaAs solar cell. We believe this continuum tailing states superposed with QD and WL induced confined energy states within the forbidden band function as an efficient carrier relaxation and collection pathway[6–8]. A slight enhancement of Jsc, accompanied by a degradation of Voc[6–12], under one sun illumination and room temperature conditions can be *completely* explained by a single-photon absorption process via below-bandgap states including both quantum confined states and tailing states. The degradation of Voc is related to an increased dark current resulting from a direct carrier collection/relaxation via the tailing states.

So far, the importance of incorporating an extended Urbach density of states has not been appreciated for the analysis of below-bandgap photocurrent generation. In the current effort of demonstrating the operation of an IBSC, some have attributed the measured photocurrent from photon excitation with energy smaller than the valance band to QD ground state transition energy to transition from QD ground state to conduction band[14]. We argue that this is not correct because this overlooks the role of an extended Urbach tail within the gap. Photons with energy lower than the valance band to quantum states transitions can still be absorbed and generate carriers in the tailing states that can be detected in the external circuit. For the demonstration of a real two-photon contributed current by absorption of two sub-bandgap photons, a quadratic increase of photocurrent with beam intensity must be observed. It is worthwhile to note that a temperature increase can also induce a non-linear photocurrent. For instance, under high intensity laser excitation, the width of the Urbach tail shows non-negligible temperature dependence. An increase in the one-photon generated current is expected with elevated temperature. This non-linear effect needs to be distinguished from a true two-photon absorption process.

The schematics of the ensemble of QDs embedded in a 2 µm i-GaAs that we used in our experiment is shown in **Figure 1(a)**. The samples that were studied were grown by molecular beam epitaxy. After depositing 2.0 ML of InAs, the InAs quantum dots with a height of 7 nm and a lateral size of ~30nm were formed with a thin wetting layer left behind. The InAs three dimensional confined islands are formed due to the lattice constant difference between InAs and GaAs and the resulting strain induced in the sample. A large spacing of 50 nm GaAs between each layer of InAs quantum dots was used in order to minimize strain accumulation and reduce the tunneling probability. A cleaved sample was fabricated with a back side n-type Au/Ge/Ni/Au electrode and a top p-type Ti/Pt/Au interdigitated electrode. As shown in our previous work, the QDs device experimentally shows a 5.5 % improvement of photo generated current density compared with its reference bulk device. While the open circuit voltage is only 0.78 V compared with 0.92 V for a bulk device. Further details can be found in one of our previous work[7].

Here we demonstrate a method to evaluate the 2PA photo-generation efficiency overcoming the limitation of traditional photocurrent spectral response measurement. Some critical advantages of our proposed method include that it is able to distinguish a two-photon absorption process from other non-linear process like local heating effect and it operates at room temperature, the most common and applicable condition used to study solar devices with unconcentrated light. In order to observe the non-linear optical transition via the extended tailing states, high intensity radiation beam with photon energy lower than the valance band to quantum dot states transition was used. A 1550 nm multi-longitudinal mode laser from Ortel Corporation amplified with an Erbium Doped Fiber Amplifier is used for the two-photon off-resonance excitation. In this experiment, the laser beam are first collimated and then focused tightly to a beam waist of order 1 µm. The output power is calibrated with an unbiased calibrated Ge detector. The quantum dot solar cell is mounted perpendicularly to the beam on a three dimensional stage with motorized actuators from Newport Corp. The photocurrent is read with a Newport 6487 Picoammeter. With the solar cell moving along the light propagation direction, the incident intensity profile changes accordingly, as illustrated in **Figure 2**, and the photocurrent dependence on intensity can be evaluated.

In the InAs/GaAs QD sample, the electrons are pumped by a sub-bandgap photon to an intermediate state (either a quantum confined state or a tailing state). Then they are extracted in the outside circuit via the tailing states or are pumped optically to the conduction band by absorbing another photon. Based on the results we previously achieved, the existence of a greatly broadened Urbach tail serves as an efficient carrier extraction pathway, which makes it possible to extract carriers created by a single photon process (1PA). If the carriers are pumped again by another photon and then collected, they contribute to the overall photon-generated current via a two-photon absorption (2PA) process. These two competing processes are illustrated in **Figure 3**.

To model the nonlinear optical interaction, the laser beam, with a Gaussian intensity profile, is described by the following expression:

$$I(r,z) = I_o \left[\frac{w_0}{w(z)}\right]^2 e^{-\frac{2r^2}{w(z)^2}} \quad (1)$$

where $r = \sqrt{x^2 + y^2}$, $w(z) = w_0\sqrt{1 + \left(\frac{z}{z_0}\right)^2}$, $I_0$ is the intensity at $z = 0$ and $r = 0$. $w_0$ is the beam waist (1.5 µm) which is measured using a knife edge method and $z_0$ is given by: $z_0 = \frac{\pi w_0^2}{\lambda} = 4.56$ µm.

The total light power is therefore $P_{total\ light\ power} = I_o \left[\frac{w_0}{w(z)}\right]^2 \int_{-\infty}^{\infty} e^{-\frac{2x^2}{w^2}} dx \int_{-\infty}^{\infty} e^{-\frac{2y^2}{w^2}} dy = \frac{\pi}{2} I_0 w_0^2$.

The 1PA is related to the available number of states and the excitation intensity. If the available state density stays constant during the measurement, the 1PA process should generate a constant number of electrons that is only dependent on the total beam power, while the 2PA process will have a quadratic dependence on light intensity. The total detected current is the sum of the single photon (1PA) and the two-photon absorption process (2PA) as expressed below:

$$i_{1PA} + i_{2PA} = i_{total} \qquad (2)$$

From an external quantum efficiency measurement (**Figure. 1(c)**) or a photoluminescence measurement (**Figure. 1(d)**), we have located the lowest energy level of the InAs/GaAs quantum dots to be around 1.12 eV. The photons with wavelength of 1550 nm excite carriers to the states below the confined quantum states. Then electrons can be collected through the continuum tailing states or can be excited again into the conduction band. The 1PA current only depends linearly on the total beam power (assuming no heating effect) and is described by the equation below:

$$i_{1PA}(z) = a \int_{-\infty}^{\infty} I(r,z) dA = a \int_{-\infty}^{\infty} I_o \left[\frac{w_0}{w(z)}\right]^2 e^{-\frac{2r^2}{w(z)^2}} 2\pi r dr =$$

$$a I_o \left[\frac{w_0}{w(z)}\right]^2 \int_{-\infty}^{\infty} e^{-\frac{2x^2}{w(z)^2}} dx \int_{-\infty}^{\infty} e^{-\frac{2y^2}{w(z)^2}} dy = \frac{\pi}{2} a I_0 w_0^2 \qquad (3)$$

The coefficient *a* here describes the *linear responsivity* of the detection process with the wavelength of 1550 nm.

Meanwhile, the 2PA process has a quadratic dependence on the beam power. It can be described by the following expression:

$$I_{2PA} = b \left[\int_{-\infty}^{\infty} I(r,z) dr\right]^2 = b \left[\int_{-\infty}^{\infty} I_o \left[\frac{w_0}{w(z)}\right]^2 e^{-\frac{4r^2}{w(z)^2}} 2\pi r dr\right]^2 = \frac{\pi}{4} b \frac{z_0^2 I_0^2 w_0^2}{z_0^2 + z^2} \qquad (4)$$

where b is a constant value indicating the efficiency of the 2PA process. The total photocurrent can be described as a sum of the 1PA and 2PA processes as:

$$I_{total} = \frac{\pi}{2} a I_0 w_0^2 + \frac{\pi}{4} b \frac{z_0^2 I_0^2 w_0^2}{z_0^2 + z^2} = \frac{\pi}{4} I_0 w_0^2 \left(2a + \frac{b I_0 z_0^2}{z_0^2 + z^2}\right) \tag{5}$$

At the beam waist, where z=0, $I_{total} = \frac{\pi}{4} I_0 w_0^2 (2a + b I_0)$.

A plot of the total measured photocurrent along the z axis under 1550 nm radiation is shown in **Figure 4(a)**, and clearly exhibits a non-linear response to the incident photon intensity. The resulting photocurrent agrees perfectly well with a Lorentzian fit in the propagation direction (as described by Equation (4)). If we plot the 2PA contributed excess photocurrent by subtracting the baseline reading from the maximum reading along with the square of the total radiation power, an exact quadratic dependence can be extracted, which is shown in **Figure 4(b)**. A slope of 0.0101 nA/mW$^2$ can be fitted, with a 2PA non-linear responsivity of 0.071 nA·μm$^2$/mW$^2$ (the value of **b**) can be extracted accordingly. From our method, the 2PA coefficient $\beta_{2PA}$ can be directly derived by relating the derived **b** value with the introduction of the responsivity factor R for 0.8 eV

$$b = \beta_{2PA} \times L \times R \tag{6}$$

where L is the effective thickness of the absorption region and R describes the amount of generated electrical charges per incident photons. Taking the photon energy as 0.8 eV, and assuming an internal quantum efficiency of 100%, R can be calculated as:

$$R = \frac{e}{2 \times 0.8 \, eV} \eta_{int} = \frac{1.6 \times 10^{-19} C}{2 \times 0.8 \times 1.6 \times 10^{-19} J} = 6.25 \times 10^5 \, \frac{nA}{mW} \tag{7}$$

Thus $\beta_{2PA}$ is extracted to be 5.7 cm/GW. It should be noted here that if the internal quantum efficiency is less than unity, the derived $\beta_{2PA}$ shall be further inversely-proportionally increased. In comparison, for bulk GaAs device, neither the 1PA photocurrent nor 2PA photocurrent is detected with our method, meaning a much smaller 1PA and 2PA coefficient within the bulk device where the Urbach energy is as small as 13 meV[1,2,6,16]. The 2PA coefficient for bulk GaAs derived by other methods such as using waveguide structure also generally shows a smaller value around 1 cm/GW[17] (we note that the assumption of this method that 2PA is the only non-linear effect will yield an overestimated value of 2PA coefficient). Therefore, we conclude that the greatly broadened tailing density of states in the QD device strongly enhances both the 1PA current generation and the two-step photon absorption process by introducing the mid-gap intermediate states.

A plot of the 1PA current vs power leads to a super-linear behavior, as shown in **Figure 4(c)**. A possible explanation for this deviation from the theoretically predicted linear dependence on total radiation power is probably due to a slight heating effect coming from the tightly focused laser beam. The Urbach tail energy ($E_U$) was found to broaden at higher temperatures[2]. When the tail extends further into the mid gap, there are more states available for low energy photon absorption. Consequentially, more generated carriers are collected via this continuum tailing density of states. The 1PA current is thus a function of the Urbach tail width. According to Fermi's Golden Rule, the transition rate increases as a result of an increase of the final available density of states. We note that in our method, the non-linear component of the 1PA current generation process can be explicitly distinguished from the 2PA process.

It is also interesting to study the X-Y dependence of the 1PA sub-bandgap photocurrent reading based on the carrier generation mechanism we proposed here. Using a laser beam focused to a beam waist of 1.5 um, the photocurrent is collected as we move the beam waist over the sample. We estimate the number of quantum dots contained in the beam to be around $1 \times 10^5$ dots. The 1PA reading is mostly constant when the beam moves in the X-Y plane, away from the edge of the sample. Even with uniform film quality like the devices we studied here, when the beam moves towards the edge of the device, the devices exhibit a slight enhancement of the base line reading when the beam is at the edge of the device (from 22 nA to 24 nA under an input power of 46.8 mW). This is very interesting because this technique can potentially be employed to scan the device area and generate a two-dimensional map as an indication of the Urbach energy variation (strain variation) across the device's whole active area with a spatial resolution only limited by the laser-beam radius. The derivation of the width of bandtail potentially can be used to predict the Voc degradation.

To conclude, the results presented here demonstrate that the extended tailing density of states in self-assembled QD solar cell can function as extra energy states for the absorption of low-energy photons and a very efficient carrier extraction pathway. In particular, a 1550 nm laser has been used to demonstrate that photons with energy smaller than the valance band to quantum dot states transition energy can facilitate both 1PA and 2PA process via the extended tailing states. Our method can clearly distinguish the photocurrent that is generated by 1PA process or a sequential absorption of two low-energy-photons. The super-linearity of the 1PA signal is potentially caused by local heating.

We gratefully acknowledge the use of the FabLab and the NISPLab facility in the University of Maryland Nanocenter.

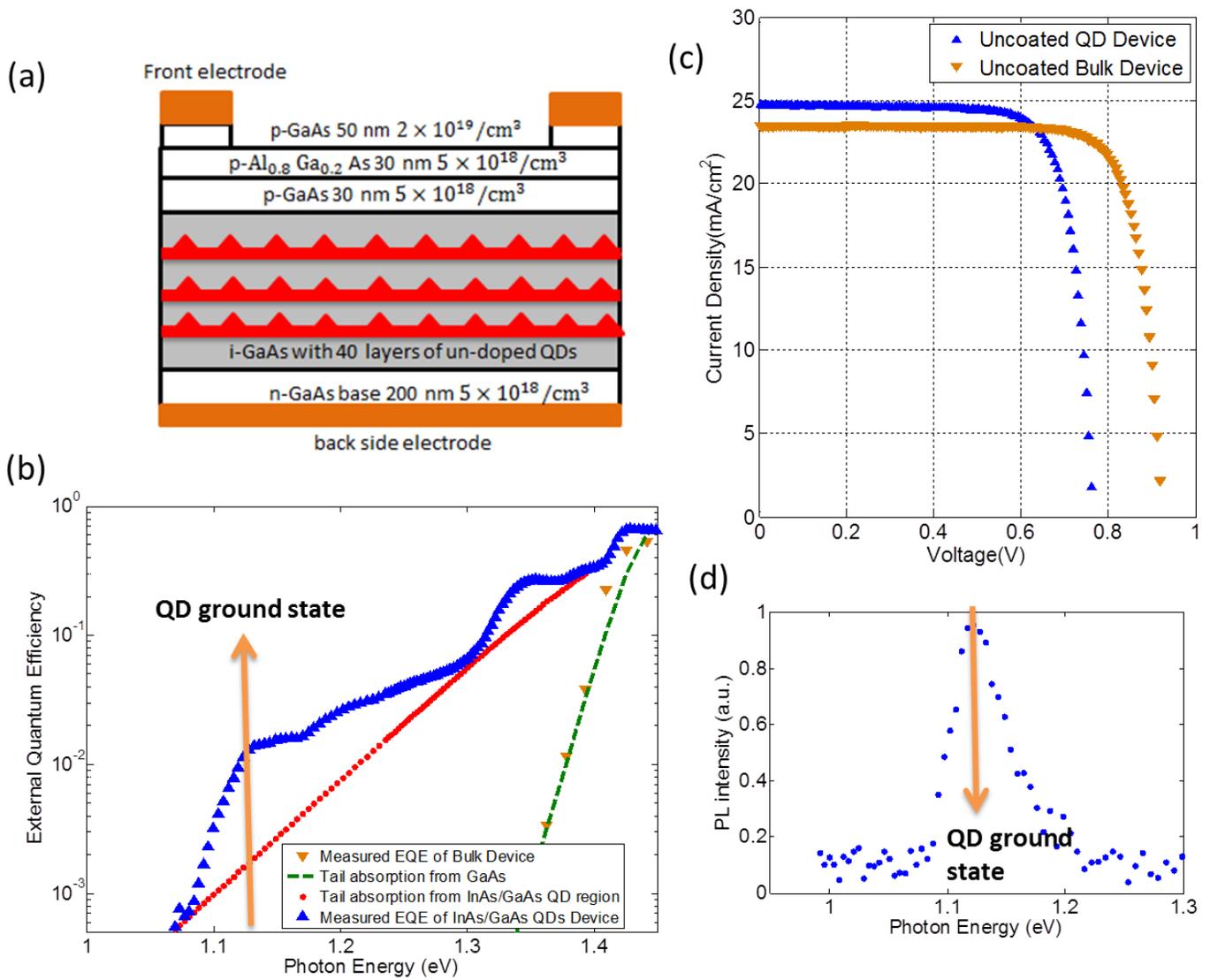

**Figure 1.** (a) Diagram of quantum dots solar cell (b) External quantum efficiency measurements and the position of tailing states of both the QD device and the bulk GaAs device (c) Current-voltage characteristics and (d) Photoluminescence measurement showing the QD ground state centered around 1.12 eV.

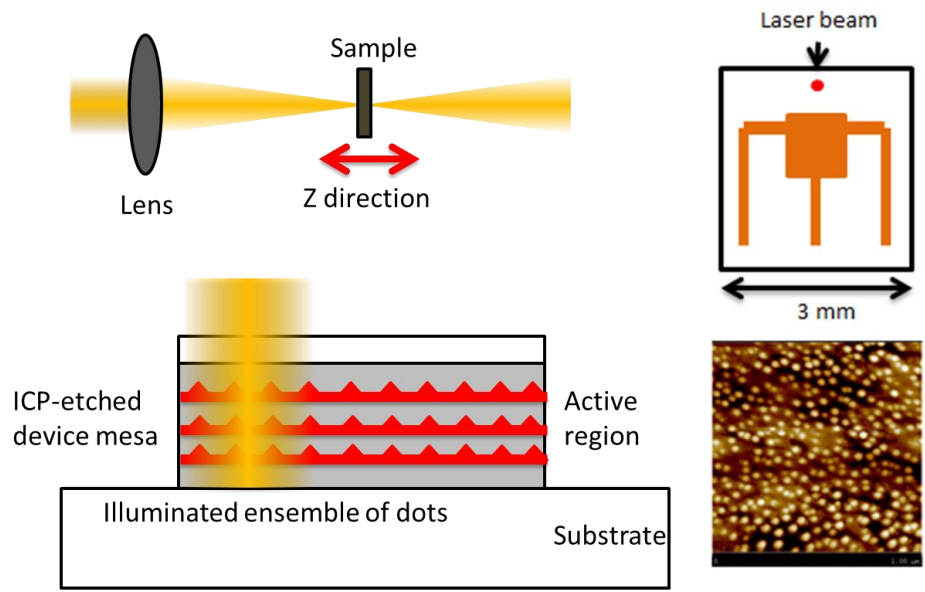

**Figure 2.** Z-scan configuration of fabricated quantum dot device

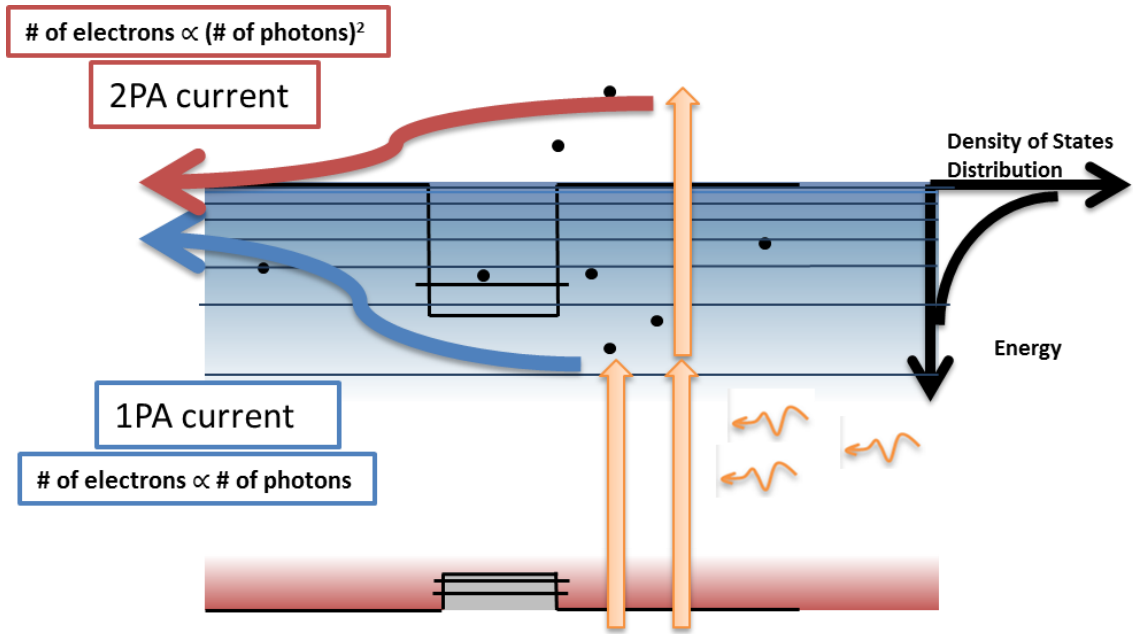

**Figure 3.** One low energy photon excites electron from the valance band to the below-bandgap tailing density of states while the tail also functions as a collection pathway so that the carriers that are generated by this 1PA process can be detected. Meanwhile, the tailing density of states can also act as an intermediate level to facilitate a two photon absorption (2PA) process. The electrons that are excited by a 2PA process will thermalize to the bandedge and will be detected as the 2PA contributed photocurrent.

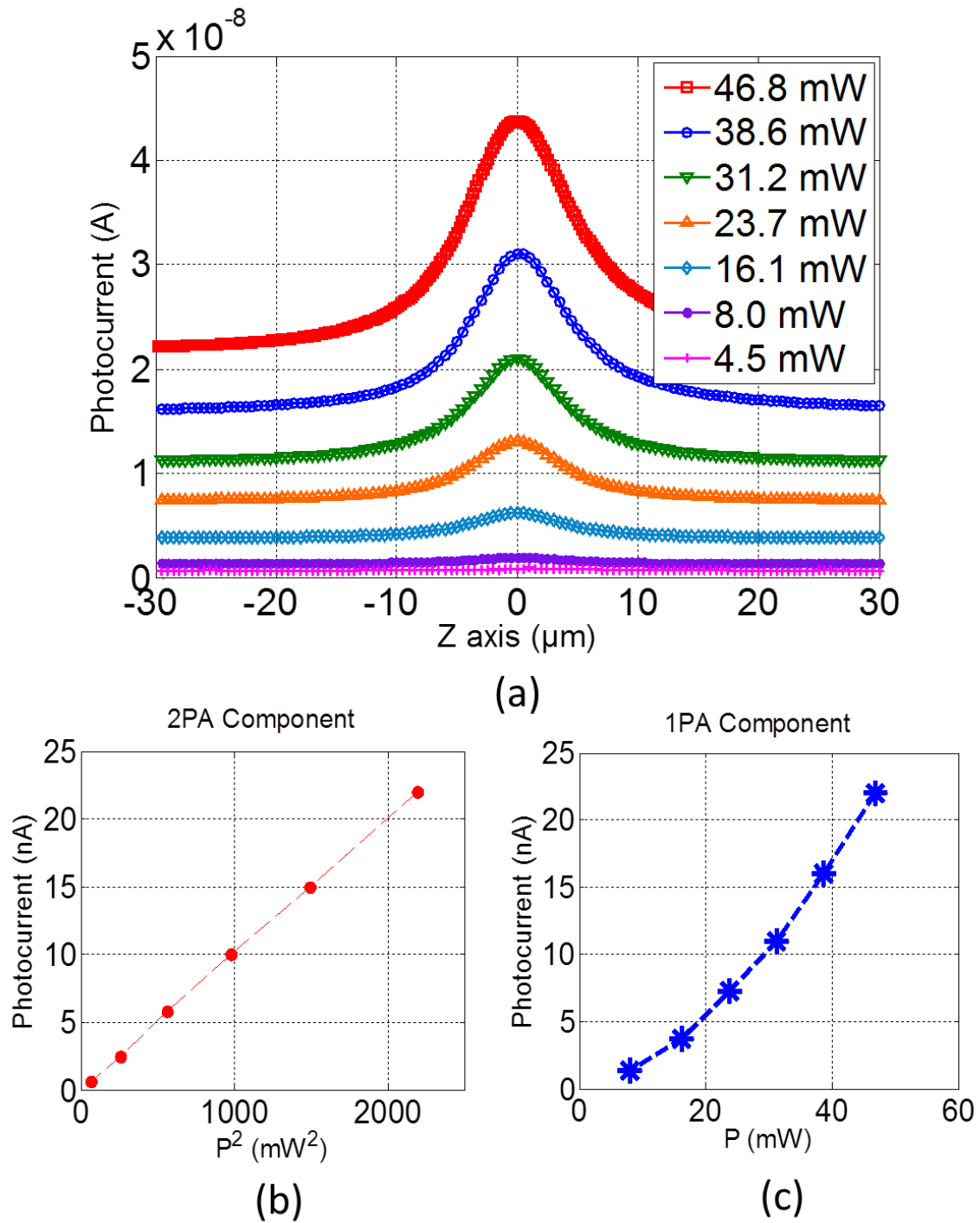

**Figure 4.** (a) Photocurrent dependence along with propagation direction for 1550 nm laser (b) The quadratic photocurrent plot indicating a 2PA photocurrent generation efficiency of around 0.01 nA/mW$^2$ (c) 1PA photocurrent generation efficiency plot with different beam power where the nonlinear component is due to local heating.